\documentclass[aps,prl,twocolumn,showpacs]{revtex4}

\usepackage{epsf}
\usepackage{bm}

\begin{document}

\title{Relative stability of excitonic complexes in quasi-one-dimensional semiconductors}

\author{I.V.~Bondarev}\email[E-mail: ]{ibondarev@nccu.edu}\affiliation{Math \& Physics Department,
North Carolina Central University, 1801 Fayetteville Str, Durham,
NC 27707, USA}

\begin{abstract}
A configuration space approach is developed to uncover generic stability peculiarities for the lowest energy neutral and charged exciton complexes (biexciton and trion) in quasi-one-dimensional semiconductors. Trions are shown to be more stable than biexcitons in strongly confined structures with small reduced electron-hole masses. Biexcitons are more stable in less confined structures with large reduced electron-hole masses. In semiconducting carbon nanotubes, in particular, the trion binding energy is shown to be greater than that of the biexciton by a factor $\sim\!1.4$ decreasing with diameter, thus revealing the general physical principles that underlie recent experimental observations.
\end{abstract}
\pacs{78.40.Ri, 73.22.-f, 73.63.Fg, 78.67.Ch}

\maketitle


Optical properties of low-dimensional semiconductor nanostructures originate from excitons and exciton complexes such as trions (charged excitons) and biexcitons. All of these have pronounced binding energies in nanostructures due to the confinement effect~\cite{HaugKoch,Cardona,Apphysrev}.~Optical properties of semiconducting carbon nanotubes (CNs), in particular, are largely determined by excitons~\cite{Dresselhaus07,Louie09}, and can be tuned by electrostatic doping~\cite{Steiner09,Spataru10,Mueller10}, or by means of the quantum confined Stark effect~\cite{Bondarev09PRB,Bondarev12PRB,Bondarev14PRB}. Trions and biexcitons, though not detectable in bulk materials at room temperature, play a significant role in quantum confined systems of reduced dimensionality such as quantum wells~\cite{Birkedal,Singh,Thilagam,Lozovik,Bracker}, nanowires~\cite{Forchel,Crottini,Sidor,Gonzales,Schuetz}, nanotubes~\cite{Pedersen03,Pedersen05,Kammerlander07,Ronnow10,Ronnow11,Matsunaga11,Santos11,Bondarev11PRB,Watanabe12,Ronnow12,Colombier12,Yuma13}, and quantum dots~\cite{Woggon,JonFbiexc,JonFtrion}.

For conventional semiconductor quantum wells, wires and dots, the binding energies of negatively or positively charged trions are known to be typically lower than those of biexcitons in the same nanostructure, although the specific trion to biexciton binding energy ratios are strongly sample fabrication dependent~\cite{Lozovik,Forchel,Sidor,Woggon}. First experimental evidence for the trion formation in carbon nanotubes was reported by Matsunaga et al.~\cite{Matsunaga11} and by Santos et al.~\cite{Santos11} on $p$-doped (7,5) and undoped (6,5) CNs, respectively. Theoretically, R{\o}nnow et al.~\cite{Ronnow10} have predicted that lowest energy trion states in all semiconducting CNs with diameters of the order of or less than 1~nm should be stable at room temperature. They have later developed the fractional dimension approach to simulate binding energies of trions and biexcitons in quasi-1D/2D semiconductors, including nanotubes as a particular case~\cite{Ronnow11,Ronnow12}. Binding energies of $63$~meV and $92$~meV are reported for the lowest energy trions~\cite{Ronnow11} and biexcitons~\cite{Ronnow12}, respectively, in the (7,5) nanotube.

However, the latest nonlinear optics experiments were able to resolve both trions and biexcitons in the same CN sample~\cite{Yuma13,Colombier12}, to report on the opposite tendency where the trion binding energy exceeds that of the biexciton rather significantly in small diameter ($\lesssim\!1$~nm) CNs. Specifically, Colombier et al.~\cite{Colombier12} reported on the observation of the binding energies $150$~meV and $106$~meV for the trion and biexciton, respectively, in the (9,7) CN. Yuma et al.~\cite{Yuma13} reported even greater binding energies of $190$~meV for the trion versus $130$~meV for the biexciton in the smaller diameter (6,5) CN. In both cases, the trion-to-biexciton binding energy ratio is greater than unity, decreasing with the CN diameter increase [1.46 for the 0.75~nm diameter (6,5) CN versus 1.42 for the 1.09~nm diameter (9,7) CN]. Trion binding energies greater than those of biexcitons are theoretically reported by Watanabe and Asano~\cite{Watanabe12}, due to the Coulomb screening effect that reduces the biexciton binding energy more than that of the trion. However, the difference calculated is at least three times less than that measured experimentally.

In this Letter, the configuration space approach first implemented in Ref.\cite{Bondarev11PRB} to evaluate biexciton binding energies in small diameter CNs, is developed to obtain the universal asymptotic relations for the lowest energy trion and biexciton binding energies in quasi-1D semiconductors. The model operates in terms of the under-barrier tunneling current between the equivalent configurations of the system in the configuration space and, therefore, allows for clear theoretical interpretation to uncover generic relative stability features of biexcitons and trions in quasi-1D semiconductors.~More specifically, whether the trion or biexciton is more stable (has greater binding energy) in a particular quasi-1D system turns out to depend on the reduced electron-hole mass and on the characteristic transverse size of the system. Trions are generally more stable than biexcitons in strongly confined quasi-1D structures with small reduced electron-hole masses, while biexcitons are more stable than trions in less confined quasi-1D structures with large reduced electron-hole masses.~For semiconducting CNs with diameters $\lesssim1\!$~nm, in particular, the model predicts the trion binding energy greater than that of the biexciton by a factor $\sim\!1.4$, decreasing with the CN diameter, in reasonable agreement with the recent experiments~\cite{Colombier12,Yuma13}. The approach was originally pioneered by Landau~\cite{LandauQM}, Gor'kov and Pitaevski~\cite{Pitaevski63}, Holstein and Herring~\cite{Herring} in the studies of molecular binding and magnetism.

\begin{figure}[t]
\epsfxsize=7.5cm\centering{\epsfbox{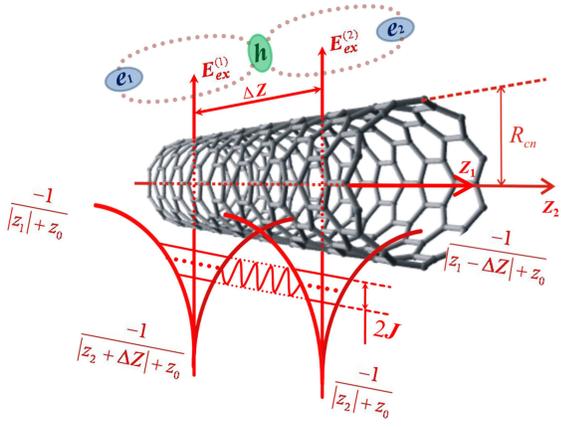}}\caption{(Color online) Schematic of the two ground-state 1D excitons sharing the same hole to form a negative trion state (arb.~units). Two collinear axes, $z_1$ and $z_2$, representing independent relative electron-hole motions in the 1st and 2nd exciton, have their origins shifted by $\Delta Z$, the inter-exciton center-of-mass distance.}\label{fig1}
\end{figure}

The problem is initially formulated for two interacting ground-state 1D excitons in a semiconducting carbon nanotube. The latter is taken as a model for definiteness. The theory and conclusions are valid for any quasi-1D semiconductor system in general. Using the cylindrical coordinate system with the \emph{z}-axis along the CN axis, as in Fig.~\ref{fig1}~(a), and separating out circumferential and longitudinal degrees of freedom of each of the excitons by transforming their longitudinal motion into their respective center-of-mass coordinates~\cite{Bondarev09PRB,Ogawa91}, one arrives at the Hamiltonian of the form~\cite{Bondarev11PRB}
\begin{eqnarray}
\hat{H}(z_1,z_2,\Delta Z)=-\frac{\partial^2}{\partial\,\!z_{1}^2}-\frac{\partial^2}{\partial\,\!z_{2}^2}\hskip2cm\label{biexcham}\\
-\frac{1}{|z_{1}|\!+\!z_0}-\!\frac{1}{|z_{1}\!-\!\Delta Z|\!+\!z_0}-\!\frac{1}{|z_{2}|\!+\!z_0}-\!\frac{1}{|z_{2}\!+\!\Delta Z|\!+\!z_0}\hskip0.3cm\nonumber\\
-\frac{2}{|(\sigma z_1+z_2)/\lambda+\Delta Z|\!+\!z_0}-\frac{2}{|(z_1+\sigma z_2)/\lambda-\Delta Z|\!+\!z_0}\nonumber\\
+\frac{2}{|\sigma(z_1-z_2)/\lambda+\Delta Z|\!+\!z_0}+\frac{2}{|(z_1-z_2)/\lambda-\Delta Z|\!+\!z_0}\;.\nonumber
\end{eqnarray}
Here, $z_{1,2}\!=\!z_{e1,2}-z_{h1,2}$ are the relative electron-hole motion coordinates of the two 1D excitons separated by the center-of-mass-to-center-of-mass distance $\Delta Z\!=\!Z_2-Z_1$, $z_0$ is the cut-off parameter of the effective (cusp-type) longitudinal electron-hole Coulomb potential, $\sigma\!=\!m_e/m_h$, $\lambda\!=\!1+\sigma$ with $m_e$ ($m_h$) representing the electron (hole) effective mass. The "atomic units"\space are used~\cite{LandauQM,Pitaevski63,Herring}, whereby distance and energy are measured in units of the exciton Bohr radius $a^\ast_B\!=\!0.529\,\mbox{\AA}\,\varepsilon/\mu$ and the Rydberg energy $Ry^\ast\!=\hbar^2/(2\mu\,m_0a_B^{\ast2})\!=\!13.6\,\mbox{eV}\,\mu/\varepsilon^2$, respectively, $\mu\!=\!m_e/(\lambda\,m_0)$ is the exciton reduced mass (in units of the free electron mass $m_0$) and $\varepsilon$ is the static dielectric constant of the electron-hole Coulomb potential.

Hamiltonian (\ref{biexcham}) is effectively two dimensional in the configuration space of the two \emph{independent} relative motion coordinates, $z_1$ and $z_2$~\cite{Endnote1}.~First two lines in Eq.~(\ref{biexcham}) represent two non-interacting 1D excitons. Their individual potentials are symmetrized to account for the presence of the neighbor a distance~$\Delta Z$ away, as seen from the $z_1$- and $z_2$-coordinate systems treated independently (Fig.~\ref{fig1}). In the $(z_1,z_2)$ configuration space the potential energy surface [second line of Eq.~(\ref{biexcham})] has four symmetrical minima to represent isolated two-exciton states~\cite{Bondarev11PRB}, separated by potential barriers responsible for the tunnel exchange coupling between these two-exciton states. Last two lines are the inter-exciton exchange Coulomb interactions --- electron-hole (line next to last) and hole-hole + electron-electron (last line), respectively.

Biexciton binding energy is $E_{X\!X}=E_g-2E_X$, where $E_g$ is the lowest eigenvalue of Eq.~(\ref{biexcham}), $E_X=-Ry^\ast/\nu_0^2$ is the single exciton binding energy with $\nu_0$ being the 1D exciton lowest-bound-state quantum number~\cite{Ogawa91}. Negative $E_{X\!X}$ indicates that the biexciton is stable with respect to the dissociation into two isolated excitons. Specifically, the tunnel exchange splitting calculation done in Ref.~\cite{Bondarev11PRB} results in
\begin{equation}
E_{X\!X}=-\frac{1}{9}\;|E_X|\left(\frac{e}{3}\right)^{2\sqrt{Ry^\ast/|E_X|}\,-\,1}\!\!\!.
\label{Exx}
\end{equation}

Trion binding energy can be found in the same way using a modification of the Hamiltonian (\ref{biexcham}), in which two same-sign particles share the third particle of an opposite sign to form the two equivalent 1D excitons as Fig.~\ref{fig1} shows for the negative trion complex consisting of the hole shared by the two electrons. Hamiltonian modified to reflect this fact has the first two lines exactly the same as in Eq.~(\ref{biexcham}), no line next to last, and one of the two terms in the last line --- either the first or the second one for the positive (with $z_{1,2}\!=\!z_{e}-z_{h1,2}$) and negative (with $z_{1,2}\!=\!z_{e1,2}-z_h$) trion, respectively. Obviously, due to the additional mass factor $\sigma$ (typically less than one for bulk semiconductors) in the hole-hole interaction term in the last line, the positive trion might be expected to have a greater binding energy in this model, in agreement with the results reported earlier~\cite{Ronnow10,Sidor}. However, the strong transverse confinement in reduced dimensionality semiconductors is known to result in the mass reversal effect~\cite{Cardona,HaugKoch}, whereby the bulk heavy hole state, the one forming the lowest excitation energy exciton of interest here, acquires a longitudinal mass comparable to the bulk \emph{light} hole mass ($\approx\!m_e$). Therefore, $m_h\!\approx\!m_e$ in our case, which is also true for CNs~\cite{Jorio05}, and so $\sigma\!=\!1$ is assumed in what follows with no substantial loss of generality. The positive-negative trion binding energy difference disappears then. The negative trion case, illustrated in Fig.~\ref{fig1}, is addressed below.

Coordinate transformation $x=(z_1-z_2-\Delta Z)/\sqrt{2}$, $y=(z_1+z_2)/\sqrt{2}$ of the original $(z_1,z_2)$ configuration space places the origin of the new coordinate system into the intersection of the two tunnel channels between the respective potential minima~\cite{Bondarev11PRB}, whereby the exchange splitting formula of Refs.~\cite{LandauQM,Pitaevski63,Herring} takes the form
\begin{equation}
E_{g,u}(\Delta Z)-2E_X=\mp J(\Delta Z)\;, \label{Egu}
\end{equation}
where
\begin{equation}
J(\Delta Z)=\int_{\!-\Delta Z/\!\sqrt{2}}^{\Delta Z/\!\sqrt{2}}\!dy\left|\psi(x,y)\frac{\partial\psi(x,y)}{\partial x}\right|_{x=0}, \label{J}
\end{equation}
$E_{g,u}$ are the ground/excited-state energies and $\psi(x,y)$ is the ground-state wave function of the Schr\"{o}dinger equation with the Hamiltonian (\ref{biexcham}) modified to the negative trion case, as discussed above, and then transformed to the $(x,y)$ coordinates. Tunnel exchange current integral $J(\Delta Z)$ is due to the electron position exchange relative to the hole (see Fig.~\ref{fig1}). This corresponds to the tunneling of the entire three particle system between the two equivalent indistinguishable configurations of the two excitons sharing the same hole in the configuration space $(z_1,z_2)$, given by the pair of minima at $z_1\!=\!z_2\!=\!0$ and $z_1\!=\!-z_2\!=\!\Delta Z$ (Fig.~\ref{fig1}). Such a tunnel exchange interaction is responsible for the coupling of the three particle system to form a stable trion state.

Function $\psi(x,y)$ in Eq.~(\ref{J}) is sought in the form~\cite{Bondarev11PRB}
\begin{equation}
\psi(x,y)=\psi_0(x,y)\exp[-S(x,y)]\,,
\label{psixy}
\end{equation}
where $\psi_0=\nu_0^{-1}\exp[-(|z_1(x,y,\Delta Z)|+|z_2(x,y,\Delta Z)|)/\nu_0]$ is the product of two single-exciton wave functions representing the isolated two-exciton state centered at the minimum $z_1\!=\!z_2\!=\!0$ (or $x\!=\!-\Delta Z/\sqrt{2}$, $y\!=\!0$) of the configuration space potential (Fig.~\ref{fig1}), and $S(x,y)$ is a \emph{slowly} varying function to take into account the deviation of $\psi$ from $\psi_0$ due to the tunnel exchange coupling to another equivalent isolated two-exciton state centered at $z_1\!=\Delta Z$, $z_2\!=\!-\Delta Z$ (or $x\!=\!\Delta Z/\sqrt{2}$, $y\!=\!0$). Substituting Eq.~(\ref{psixy}) into the Schr\"{o}dinger equation with the negative trion Hamiltonian pre-transformed to the $(x,y)$ coordinates, one obtains in the region of interest
\[
\frac{\partial S}{\partial x}=\frac{\nu_0}{\Delta Z/\sqrt{2}-x}
\]
($|x|\!<\!\Delta Z/\sqrt{2}$, cut-off $z_0$ dropped~\cite{Bondarev11PRB}) up to negligible terms of the order of the second derivatives of~$S$.~This equation is to be solved with the boundary condition $S(-\Delta Z/\sqrt{2},y)\!=\!0$ originating from the natural requirement $\psi(-\Delta Z/\sqrt{2},y)\!=\!\psi_0(-\Delta Z/\sqrt{2},y)$, to result in
\begin{equation}
S(x,y)=\nu_0\ln\frac{\sqrt{2}\Delta Z}{\Delta Z/\sqrt{2}-x}.
\label{sxy}
\end{equation}

\begin{figure}[t]
\epsfxsize=8.65cm\centering{\epsfbox{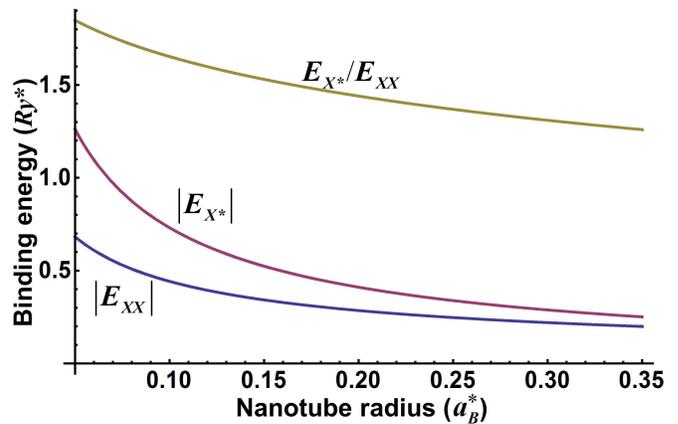}}\caption{(Color online) Trion binding energy, biexciton binding energy and their ratio given by Eqs.~(\ref{Exstar}), (\ref{Exx}) and (\ref{ExstarExx}), respectively, with $|E_X|\!=\!Ry^\ast\!/r^{0.6}$ (Ref.~\cite{Pedersen03}) as functions of the dimensionless nano\-tube radius.}\label{fig2}
\end{figure}

\begin{figure}[b]
\epsfxsize=8.65cm\centering{\epsfbox{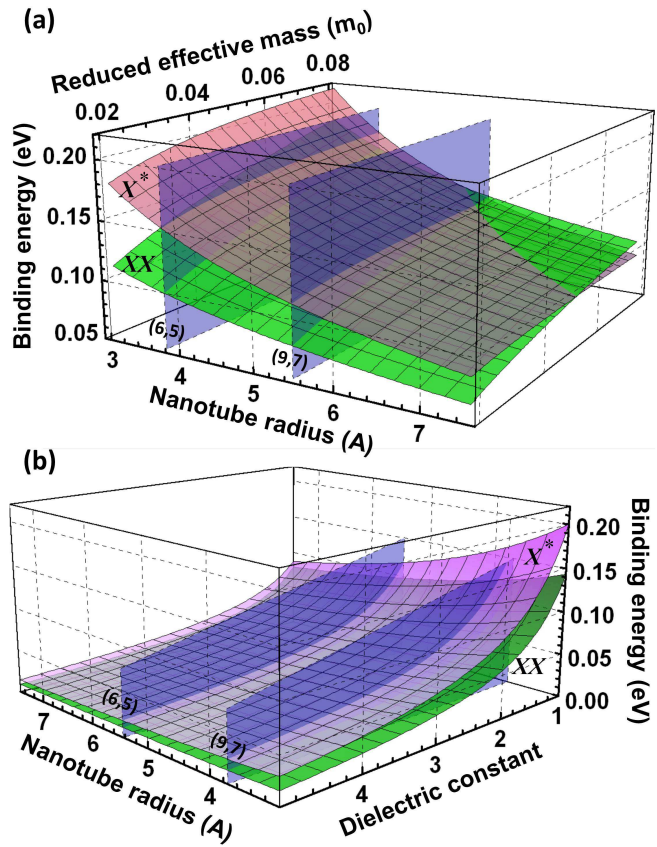}}\caption{(Color online) Trion ($X^\ast$) and biexciton ($X\!X$) binding energies given by Eqs.~(\ref{Exstar}) and (\ref{Exx}) with $|E_X|\!=\!Ry^\ast\!/r^{0.6}$, as functions of the CN radius and $\mu$ with $\varepsilon\!=\!1$ (a), and as functions of the CN radius and $\varepsilon$ with $\mu\!=\!0.04$ (b). Vertical parallel planes indicate the radii of the (6,5) and (9,7) CNs studied experimentally.}\label{fig3}
\end{figure}

After plugging Eqs.~(\ref{sxy}) and (\ref{psixy}) into Eq.~(\ref{J}), and retaining only the leading term of the integral series expansion in powers of $\nu_0$ subject to $\Delta Z>1$, one obtains
\begin{equation}
J(\Delta Z)=\frac{2}{2^{2\nu_0}\nu_0^3}\Delta Z\,e^{-2\Delta Z/\nu_0}. \label{Jfin}
\end{equation}
The ground state energy $E_g$ of the three particle system in Eq.~(\ref{Egu}) is now seen to go through the negative minimum (the trion state) as $\Delta Z$ increases. The minimum occurs at $\Delta Z_0=\nu_0/2$, whereby the trion binding energy~is $E_{X^{\!\ast}}\!=\!-J(\Delta Z_0)\!=\!-1/(e\,2^{2\nu_0}\nu_0^2)$. In absolute units, expressing $\nu_0$ in terms of $E_X$, one has
\begin{equation}
E_{X^{\!\ast}}=-\frac{|E_X|}{e\,2^{2\sqrt{Ry^\ast/|E_X|}}}
\label{Exstar}
\end{equation}
with the trion-to-biexciton binding energy ratio
\begin{equation}
\frac{E_{X^{\!\ast}}}{E_{X\!X}}=3\!\left(\frac{3}{2e}\right)^{\!2\sqrt{Ry^\ast/|E_X|}}\!\!,
\label{ExstarExx}
\end{equation}
according to Eq.~(\ref{Exx}).

Now assuming $|E_X|\!=\!Ry^\ast\!/r^{0.6}$ with $r$ being the dimensionless CN radius, as was reported by Pedersen from variational calculations~\cite{Pedersen03}, one has the $r$-dependences of $|E_{X^{\!\ast}}|$, $|E_{X\!X}|$ and $E_{X^{\!\ast}}\!/E_{X\!X}$ as plotted in Fig.~\ref{fig2}.~The trion and biexciton binding energies both decrease with increasing $r$ --- in such a way that their ratio remains greater than unity for small enough $r$ --- in full agreement with the experiments by Colombier et al.~\cite{Colombier12} and Yuma et al.~\cite{Yuma13}. However, since the factor $3/2e$ in Eq.~(\ref{ExstarExx}) is less than one, the ratio can also be less than unity for $r$ large enough, but not too large, so that the 1D model used here still works. Interestingly, as $r$ tends to zero, Eq.~(\ref{ExstarExx}) yields $E_{X^{\!\ast}}\!/E_{X\!X}\approx3$ as the pure 1D limit for the trion-to-biexciton binding energy ratio. When $E_{X^{\!\ast}}\!/E_{X\!X}$ is known, one can use Eq.~(\ref{ExstarExx}) to estimate the effective Bohr radii $a^\ast_B$ for the excitons in the CNs of known radii. For example, substituting $1.46$ for the 0.75~nm diameter (6,5) CN and $1.42$ for the 1.09~nm diameter (9,7) CN, as reported by Yuma et al.~\cite{Yuma13} and Colombier et al.~\cite{Colombier12}, respectively, into the left hand side of the transcendental equation~(\ref{ExstarExx}) and solving it for $a^\ast_B$, one obtains the effective exciton Bohr radius $2$~nm and $2.5$~nm for the (6,5) CN and (9,7) CN, respectively, in reasonable agreement with previous estimates~\cite{Pedersen03,Yuma13}.

\begin{figure}[t]
\epsfxsize=8.65cm\centering{\epsfbox{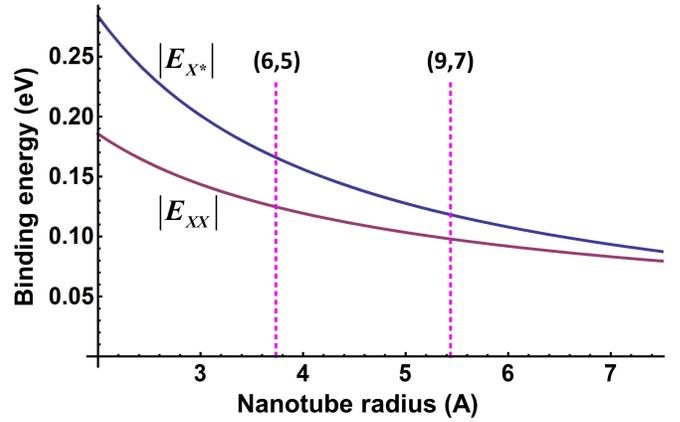}}\caption{(Color online) Cross-section of Fig.~\ref{fig3}(a) at $\mu\!=\!0.04$ showing the relative behavior of the trion and biexciton binding energies in semiconducting CNs of increasing radius.}\label{fig4}
\end{figure}

In general, the binding energies in Eqs.~(\ref{Exstar}) and (\ref{Exx}) are functions of the CN radius, $\mu$ and~$\varepsilon$. Figures~\ref{fig3}~(a) and (b) show their 3D plots at fixed $\varepsilon\;(=\!1)$ and $\mu\;(=\!0.04)$, respectively, as functions of two remaining variables. The reduced effective mass $\mu$ chosen is typical of large radius excitons in small-diameter CNs~\cite{Jorio05}.~The unit dielectric constant $\varepsilon$ assumes the CN placed in air and the fact that there is no screening in quasi-1D semiconductor systems both at short and at large electron-hole separations~\cite{Louie09}. This latter assumption of the unit background dielectric constant remains legitimate for \emph{small} diameter ($\lesssim\!1$~nm) semiconducting CNs in dielectric screening environment, too, --- for the lowest excitation energy exciton in its ground state of interest here (not for its excited states though), in which case the environment screening effect is shown by Ando to be negligible~\cite{Ando2010}, diminishing quickly with the increase of the effective distance between the CN and dielectric medium relative to the CN diameter.

Figure~\ref{fig3}~(a) can be used to evaluate the relative stability of the trion and biexciton complexes in quasi-1D semiconductors.~We see that whether the trion or the biexciton is more stable in a particular quasi-1D system depends on $\mu$ and on the characteristic transverse size of the nanostructure.~In strongly confined quasi-1D systems with relatively small $\mu$, such as small-diameter CNs, the trion is generally more stable than the biexciton. In less confined quasi-1D structures with greater $\mu$ typical of semiconductors~\cite{Cardona}, the biexciton is more stable than the trion.~This is a generic peculiarity in the sense that it comes from the tunnel exchange in the quasi-1D electron-hole system in the configuration space. Greater $\mu$, while not affecting significantly the single charge tunnel exchange in the trion complex, makes the neutral biexciton complex generally more compact, facilitating the mixed charge tunnel exchange in it and thus increasing the stability of the complex. From Fig.~\ref{fig3}~(b) we see that this generic feature is not affected by the variation of $\varepsilon$, although the increase of $\varepsilon$ decreases the binding energies of both excitonic complexes --- in agreement both with theoretical studies~\cite{Ronnow10} and with experimental observations of lower binding energies (compared to those in CNs) of these complexes in conventional semiconductor nanowires~\cite{Forchel,Crottini,Sidor,Gonzales,Schuetz}. The latter are self-assembled nano\-structures of one (transversely confined) semiconductor embedded in another (bulk) semiconductor with the characteristic transverse confinement size typically greater than that of small diameter CNs, and so both inside and outside material dielectric properties matter.

Figure~\ref{fig4} shows the cross-section of Fig.~\ref{fig3}~(a) taken~at $\mu\!=\!0.04$ to present the relative behavior of $|E_{X^{\!\ast}}|$ and $|E_{X\!X}|$ in semiconducting CNs of increasing radius. Both $|E_{X^{\!\ast}}|$ and $|E_{X\!X}|$ decrease, and so does their ratio, as the CN radius increases. From the graph, $|E_{X^{\!\ast}}|\!\approx\!170$ and $125$~meV, $|E_{X\!X}|\!\approx\!120$ and $95$~meV, for the (6,5) and (9,7) CNs, respectively. This is to be compared with $190$ and $130$~meV for the (6,5) CN~\cite{Yuma13} versus $150$ and $106$~meV for the (9,7) CN~\cite{Colombier12} reported experimentally. We see that, as opposed to perturbative theories~\cite{Watanabe12}, the present theory underestimates experimental data just slightly, most likely due to variational treatment limitations. It does explain well the trends observed, and so the graph in Fig.~\ref{fig4} can be used as a guide for trion and biexciton binding energy estimates in small diameter ($\lesssim\!1$~nm) nanotubes.

To summarize, presented  herein are the generic stability features for neutral and charged exciton complexes in quasi-1D semiconductors.~Trions are shown to be more stable than biexcitons in strongly confined quasi-1D structures with small reduced electron-hole masses. Biexcitons are more stable in less confined structures with large reduced electron-hole masses.~In a particular case of small diameter semiconducting CNs, the calculated trion binding energy is greater than that of the biexciton by a factor $\sim\!1.4$, decreasing with the CN diameter, thus revealing the general physical principles that underlie recent experimental observations~\cite{Colombier12,Yuma13}.

This work is supported by DOE (DE-SC0007117). I.V.B. acknowledges discussion with D.Tomanek of MSU and thanks T.Heinz of Columbia U. for pointing out Ref.\cite{Louie09} of relevance to this work.


\begin{thebibliography}{[99]}
\bibitem{Endnote1}The intra-exciton motion is assumed to be faster than the inter-exciton center-of-mass relative motion, which is always the case as the exciton itself is normally more stable than its compound complexes, thus justifying the adiabatic approximation used in Eq.~(\ref{biexcham}).
\end{thebibliography}
\end{document}